\documentclass[10pt]{article}

\usepackage{multirow}
\usepackage{booktabs}
\usepackage{tabularx}
\usepackage{ltablex}
\usepackage{array}

\usepackage{amssymb,amsmath,amsthm}
\usepackage{nicefrac,bbm}
\theoremstyle{definition}   
\usepackage{apptools,mathtools,chngcntr}

\usepackage{color,xcolor}
\usepackage[dvipsnames]{xcolor}
\usepackage{tikz,pgfplots}
\usepgfplotslibrary{colormaps,fillbetween,groupplots}
\usepackage{pgfplotstable}
\usepackage{subcaption}
\usepackage[font=small]{caption}
\usepackage{wrapfig}
\pgfplotsset{compat=newest}
\usetikzlibrary{patterns,shapes,arrows.meta,patterns.meta,tikzmark,matrix,calc}
\tikzset{>=latex} 
\usepackage{pgfgantt}

\usepackage[maxbibnames=50, maxcitenames=2, uniquelist=false,
    natbib=true, style=authoryear-comp,
    isbn=false, url=false,giveninits=true,dashed=false]{biblatex}   
\bibliography{bibliography}
\AtBeginBibliography{} 

\usepackage[utf8]{inputenc}
\usepackage[T1]{fontenc}
\usepackage[left=20mm,right=20mm,top=20mm,bottom=25mm,a4paper]{geometry}
\usepackage[hidelinks]{hyperref}
\usepackage{pdfpages}
\usepackage[capitalise]{cleveref}
\crefname{section}{Section}{Sections}
\crefname{appendix}{Appendix}{Appendices}
\crefname{table}{Table}{Tables}
\usepackage{titlesec, blindtext}
\usepackage{marginnote}

\newtheorem{definition}{Definition}[section]
\newtheorem{remark}[definition]{Remark}

\usepackage[affil-it]{authblk}
\newcommand{\ba}{\boldsymbol{a}}
\newcommand{\bb}{\boldsymbol{b}}

\newcommand{\bn}{\boldsymbol{n}}
\newcommand{\bp}{\boldsymbol{k}}
\newcommand{\bw}{\boldsymbol{w}}
\newcommand{\bx}{\boldsymbol{x}}

\newcommand{\bA}{\boldsymbol{A}}
\newcommand{\bB}{\boldsymbol{B}}

\newcommand{\bF}{\boldsymbol{F}}
\newcommand{\bG}{\boldsymbol{G}}
\newcommand{\bH}{\boldsymbol{H}}
\newcommand{\bI}{\boldsymbol{I}}

\newcommand{\bP}{\boldsymbol{P}}

\newcommand{\bW}{\boldsymbol{W}}

\newcommand{\cA}{{\mathcal{A}}}
\newcommand{\cB}{{\mathcal{B}}}

\newcommand{\cD}{{\mathcal{D}}}

\newcommand{\cF}{{\mathcal{F}}}

\newcommand{\cI}{{\mathcal{I}}}

\newcommand{\cL}{{\mathcal{L}}}

\newcommand{\cP}{{\mathcal{P}}}

\newcommand{\cW}{{\mathcal{W}}}

\newcommand{\bbA}{{\mathbb{A}}}

\newcommand{\bbR}{{\mathbb{R}}}

\newcommand{\bnull}{\boldsymbol{0}}

\newcommand{\SP}{\mathcal{SP}}

\newcommand{\cof}{\operatorname{cof}}

\newcommand{\bcI}{\boldsymbol{\cI}}

\newcommand{\norm}[1]{\left\lVert#1\right\rVert}

\newcommand{\IoneISO}{I_1^{\text{iso}}}
\newcommand{\ItwoISO}{I_2^{\text{iso}}}

\newcommand{\IoneTI}{I_1^{\text{ti}}}
\newcommand{\ItwoTI}{I_2^{\text{ti}}}

\definecolor{CPSgreen}{RGB}{22,164,138}
\definecolor{CPSlightblue}{RGB}{104,143,198}
\definecolor{CPSdarkblue}{RGB}{67,83,132}
\definecolor{CPSgrey}{RGB}{204, 204, 204}
\definecolor{CPSorange}{RGB}{246,163,21}
\definecolor{CPSred}{RGB}{194,76,76}
\definecolor{CPSdarkgrey}{RGB}{90, 90, 90}

\definecolor{MAXgrey}{RGB}{225,225,225}

\pgfplotscreateplotcyclelist{colorlistBCCPc}{%
only marks,solid, mark options={solid,fill=CPSred,fill opacity=1,mark size=4pt},mark=square*,CPSred\\
only marks,solid, mark options={solid,fill=CPSorange,fill opacity=1,mark size=4pt},mark=square*,CPSorange\\
only marks,solid, mark options={solid,fill=CPSdarkblue,fill opacity=1,mark size=4pt},mark=square*,CPSdarkblue\\
only marks,solid, mark options={solid,fill=CPSlightblue,fill opacity=1,mark size=4pt},mark=square*,CPSlightblue\\
CPSred, dashed, very thick,mark=*, mark options={solid}, mark repeat = 15\\
CPSorange, dashed, very thick,mark=*, mark options={solid} , mark repeat = 15\\
CPSdarkblue, dashed, very thick ,mark=*, mark options={solid}, mark repeat = 15\\
CPSlightblue, dashed, very thick ,mark=*, mark options={solid}, mark repeat = 15\\
CPSred, very thick \\
CPSorange, very thick \\
CPSdarkblue, very thick \\
CPSlightblue, very thick \\
}

\pgfplotscreateplotcyclelist{mycolorlist2}{%
only marks,solid, mark options={solid,fill=color11,fill opacity=1,mark size=4pt},mark=square*,color11\\
only marks,solid, mark options={solid,fill=color12,fill opacity=1,mark size=4pt},mark=square*,color12\\
only marks,solid, mark options={solid,fill=color13,fill opacity=1,mark size=4pt},mark=square*,color13\\
only marks,solid, mark options={solid,fill=color21,fill opacity=1,mark size=4pt},mark=square*,color21\\
only marks,solid, mark options={solid,fill=color22,fill opacity=1,mark size=4pt},mark=square*,color22\\
only marks,solid, mark options={solid,fill=color23,fill opacity=1,mark size=4pt},mark=square*,color23\\
only marks,solid, mark options={solid,fill=color31,fill opacity=1,mark size=4pt},mark=square*,color31\\
only marks,solid, mark options={solid,fill=color32,fill opacity=1,mark size=4pt},mark=square*,color32\\
only marks,solid, mark options={solid,fill=color33,fill opacity=1,mark size=4pt},mark=square*,color33\\
color11 \\
color11 \\
color11, dashed, very thick,mark=*, mark options={solid}, mark repeat = 15\\
color12, dashed, very thick ,mark=*, mark options={solid}, mark repeat = 15\\
color13, dashed, very thick ,mark=*, mark options={solid}, mark repeat = 15\\
color21, dashed, very thick ,mark=*, mark options={solid}, mark repeat = 15\\
color22, dashed, very thick ,mark=*, mark options={solid}, mark repeat = 15\\
color23, dashed, very thick ,mark=*, mark options={solid}, mark repeat = 15\\
color31, dashed, very thick,mark=*, mark options={solid} , mark repeat = 15\\
color32, dashed, very thick ,mark=*, mark options={solid}, mark repeat = 15\\
color33, dashed, very thick,mark=*, mark options={solid} , mark repeat = 15\\
color11, very thick \\
color12, very thick \\
color13, very thick \\
color21, very thick \\
color22, very thick \\
color23, very thick \\
color31, very thick \\
color32, very thick \\ 
color33, very thick \\
color11 \\
color11 \\
}

\pgfplotscreateplotcyclelist{mycolorlist123}{%
only marks,solid, mark options={solid,fill=color11,fill opacity=1,mark size=4pt},mark=square*,color11\\
only marks,solid, mark options={solid,fill=color22,fill opacity=1,mark size=4pt},mark=square*,color22\\
only marks,dashed, mark options={solid,fill=color33,fill opacity=1,mark size=4pt},mark=square*,color33\\
color11, very thick,dashed, mark options={solid}, mark repeat = 15\\
color22, very thick , mark options={solid}, mark repeat = 15\\
color33, very thick,dashed , mark options={solid}, mark repeat = 15\\
}

\pgfplotscreateplotcyclelist{mycolorlistSNE}{%
CPSdarkgrey,  thick, dashed , mark options={solid}, mark repeat = 15, mark=*\\
CPSred, ultra thick \\
}

\pgfplotscreateplotcyclelist{mycolorlistTRELOAR}{%
CPSdarkblue,  thick, dashed , mark options={solid}, mark repeat = 2, mark=*\\
CPSlightblue, ultra thick \\
CPSdarkblue, ultra thick \\
}

\pgfplotscreateplotcyclelist{colorlistBCCintro}{%
CPSdarkblue,  very thick, loosely dashed , mark options={solid}, mark repeat = 15, mark=*\\
CPSdarkblue,  ultra thick \\
CPSlightblue, ultra thick \\
}

\pgfplotscreateplotcyclelist{colorlistBCCNonPc}{%
only marks,solid, mark options={solid,fill=CPSred,fill opacity=1,mark size=4pt},mark=square*,CPSred\\
only marks,solid, mark options={solid,fill=CPSorange,fill opacity=1,mark size=4pt},mark=square*,CPSorange\\
only marks,solid, mark options={solid,fill=CPSred,fill opacity=1,mark size=4pt},mark=square*,CPSred\\
only marks,solid, mark options={solid,fill=CPSorange,fill opacity=1,mark size=4pt},mark=square*,CPSorange\\
CPSorange,  very thick,   loosely dashed , mark options={solid}, mark repeat = 15, mark=*\\
CPSred,  very thick,  loosely dashed , mark options={solid}, mark repeat = 15, mark=*\\
CPSorange,  ultra thick \\
CPSred, ultra thick \\
}

\tikzstyle{connect}=[thick,black]
\tikzstyle{connect arrow}=[-{Latex[length=4,width=3.5]},thick,black,shorten <=0.5,shorten >=1]
\tikzstyle{connect arrow pub}=[-{Latex[length=4,width=3.5]}, thick,black]
\tikzstyle{connect arrow dashed}=[-{Latex[length=4,width=3.5]},ultra thick,black,shorten <=0.5,shorten >=1, 
dashed]

\tikzstyle{connect arrow dashed 2}=[connect, densely dashed, -{Latex[length=4,width=3.5]}, very thick]

\tikzstyle{connect2}=[thick, black, shorten <=1, shorten >=1, very thick] 
\tikzstyle{connect arrow2}=[connect2, -{Latex[length=4,width=3.5]}, very thick]

\tikzstyle{connect arrow constraint}=[connect arrow2, CPSred, very thick]
\tikzstyle{connect constraint}=[connect2, CPSred, very thick] 

\tikzstyle{node}        =[thick, draw=black, text = black, minimum size=25, inner sep=0.5, outer sep=0.6, rounded corners = 2]
\tikzstyle{node blue}=[node, CPSlightblue!10!black, fill=CPSlightblue!20]
\tikzstyle{node orange}=[node, CPSorange!10!black, fill=CPSorange!20,]
\tikzstyle{node red}=[node, CPSred!10!black, fill=CPSred!20]
\tikzstyle{node green}=[node, CPSgreen!10!black, fill=CPSgreen!20, dashed]

\newcommand{\layernode}[6][1]{
    \node[node 2] (#4) at (#2, #3) {};
    \ifnum#6=1 
        \draw[domain=-4:4, shift={(#2, #3-0.1)}, very thick, scale=0.05, smooth, variable=\x, black] plot ({\x}, {ln(1+exp(1.5*\x))});
    \else
        \ifnum#6=2 
            \draw[domain=-4:4, shift={(#2, #3)}, very thick, scale=0.05, smooth, variable=\x, black] plot ({\x}, {\x});
        \else
            \ifnum#6=3 
                \draw[domain=-4:4, shift={(#2, #3-0.1)}, very thick, scale=0.05, smooth, variable=\x, black] plot ({\x}, {1.5*max(\x,0)});
            \else
                \ifnum#6=5
            \draw[domain=-4:4, shift={(#2, #3)}, very thick, scale=0.05, smooth, variable=\x, black] plot ({\x}, {(0.6*\x)*(0.6*\x)-4.5});
        \else
                \draw[domain=-4:4, shift={(#2, #3)}, very thick, scale=0.05, smooth, variable=\x, black] plot ({\x}, {3*tanh(\x)});
            \fi
            \fi
        \fi
    \fi
    \
    \node[above left] at (#2, #3) {\small #5};
    \ifnum#1=1 
        \draw[connect arrow] (#2, #3+0.7) -- (#4);
    \fi
}

\title{Structured sparsification of hyperelastic\\physics-augmented neural networks
\vspace{1ex}}

\author[1]{Anne~Kadereit}
\author[2,*]{Dominik~K.~Klein}
\author[1]{Marlon~Franke}
\author[2]{Oliver Weeger}

\affil[1]{\footnotesize Institute of Mechanics, Karlsruhe Institute of Technology, 76131 Karlsruhe, Germany}
\affil[2]{\footnotesize Cyber-Physical Simulation, 
Department of Mechanical Engineering,\protect\\Technical University of Darmstadt, 64293 Darmstadt, Germany}
\affil[*]{\footnotesize Corresponding author, email: klein@cps.tu-darmstadt.de}

\date{November 13, 2026}
 
\begin{document}

\maketitle

\par\noindent\rule{\textwidth}{0.4pt}
\begin{abstract}

Physics-augmented neural network (PANN) constitutive models combine the flexibility of feed-forward neural networks (FFNNs) with a sound mechanical basis, which makes them a promising approach to representing complex material behaviors. Thereby, one challenge is the choice of network architecture, i.e., the number of inputs, hidden layers, and neurons, which is typically done based on expert knowledge or hyperparameter studies. In this work, we examine structured sparsification for hyperelastic PANN models, such that a reasonably small network size is determined during calibration. For this purpose, gates are assigned to all inputs and neurons of the FFNN, and an $L_2$ regularization is applied to the weights and gates. The approach is investigated for polyconvex and unrestricted PANN models with one and two hidden layers, calibrated to data of a transversely isotropic rank-one laminate. The final network size turns out to be largely independent of the initial network size and is mainly controlled by the penalty parameter, which governs the trade-off between network size and accuracy. For a given number of parameters, the sparsified networks are at least as accurate as networks trained directly. Thereby, compact models with only a few neurons are obtained that still represent the material behavior accurately.

\end{abstract}
\vspace*{2ex}
\par\noindent\rule{\textwidth}{0.4pt}

\section{Introduction}\label{sec:intro}

Neural constitutive models have emerged in recent years as a promising approach to representing complex material behaviors. There is broad agreement that such models should incorporate constitutive conditions to provide mechanically reasonable predictions, while exploiting the flexibility of neural networks. This is referred to as thermodynamics-based \parencite{masi2021}, mechanics-informed \parencite{ASAD2023116463}, physics-based \parencite{aldakheel2025}, physics-constrained \parencite{kalina2022b}, physics-augmented \parencite{klein2022b}, or constitutive artificial \parencite{Linka2020} neural network (NN) modeling. By now, the formulation and application of hyperelastic neural constitutive models are largely understood \parencite{fuhg2024,KLEIN2027106815,Linka2023,Geuken2026,linden2023}. Moreover, considerable progress has been made in the formulation of multiphysical \parencite{HOLTHUSEN2026119082,kalina2024a,klein2022b} and inelastic \parencite{FRIEDRICHS2026118852,boes2025accountingplasticityextensioninelastic,Rosenkranz2024,ASAD2023116463} neural constitutive models.

\medskip

One challenge in the application of neural constitutive models is the selection of suitable hyperparameters. For instance, for models based on feed-forward NNs (FFNNs), the number of hidden layers and neurons has to be chosen, as well as the inputs, e.g., which invariants are used. In practice, this can easily result in models with a large number of parameters \parencite[Tab.~2]{Fernandez2020}, \parencite[Tab.~2]{klein2022a}. Since such models are robust against overfitting when suitable mechanical conditions are included, a large number of parameters is not necessarily a problem for the model generalization \parencite{linden2023}. However, when the models are applied in finite element analysis, the computational cost can increase with the number of parameters \parencite[Fig.~8]{franke2023}, \parencite[App.~E]{ALHEIT2026118728}. Moreover, it has been reported that very compact NN architectures can be sufficiently flexible for many applications \parencite{kalina2025,klein2024a}. From a modeling point of view, it therefore seems appealing to reduce the model complexity and the number of trainable parameters whenever possible. 

\medskip

A heuristic approach to finding reasonably small network architectures is a hyperparameter study: initially, very small networks are trained, which may not be flexible enough to represent the given material behavior, and the number of trainable parameters is then increased until the material behavior is represented sufficiently well \parencite[App.~E]{kalina2025}, \parencite[Sec.~4.2]{Klein2026a}. While such hyperparameter studies may be guided by expert knowledge and literature standards, they can still be cumbersome. Alternatively, one can start from a sufficiently large network architecture and try to reduce it during the training process. This idea is well-established in machine learning, where it is frequently realized by suitable $L_p$ regularization, i.e., by penalizing the model parameters in the loss function such that some of them become exactly zero. By this, model parameters can be removed during training \parencite{kolb_smoothing_2026,louizos_learning_2018}. 

\medskip

The earliest work on sparsification of neural constitutive models that we are aware of is by~\textcite{fuhg_extreme_2024}, who employ a smoothed $L_0$ regularization on the individual model parameters \parencite{louizos_learning_2018}. Across a variety of model classes and datasets, the authors demonstrate that FFNN-based models can be sparsified to such an extent that the remaining model equations can be written down in a single line after training. Such $L_0$-based sparsification of individual model parameters has since been adopted in several works \parencite{tan_towards_2026,anantha_padmanabha_improving_2024,linka_best-in-class_2024}. Since $L_0$ penalizes the \emph{number} of non-zero model parameters but not their \emph{magnitude}, it is a natural choice for promoting sparsification. However, a loss function based on $L_0$ is not smooth, so careful numerical treatment is required to make the training process feasible. Further component-wise $L_p$ regularizations for neural constitutive models were investigated by~\textcite{mcculloch_sparse_2024}, who found that $p\leq 1$ induces sparsity, whereas, e.g., $p=2$ does not. Similar observations have been reported by \textcite{alheit_cann-euclid_2026,martonova_automated_2024,hou_exploring_2025}. Apart from reducing the model size, sparsification has also been used to gain mechanical insights into the underlying material behavior, e.g., by identifying its material symmetry \parencite{kalina2025} or degree of inelasticity \parencite{holthusen_generalized_2026}. Moreover, a partial sparsification of the model parameters has been reported even without explicit $L_p$ regularization \parencite{Linka2023}, which might be caused by non-negativity constraints on some of the model parameters \parencite{Klein2026a}. Beyond classical FFNNs, sparsification has also been employed in neural constitutive models based on Kolmogorov-Arnold networks. There, the norms of the trainable functions on the network edges are penalized, which allows to identify compact expressions for the constitutive model equations \parencite{abdolazizi_constitutive_2025,ji_inelastic_2026}. Besides sparsification of neural networks, further methods have been proposed to identify compact material models, e.g., symbolic regression \parencite{ABDUSALAMOV2024104037} and model library approaches \parencite{flaschel2021,FLASCHEL2026118573}.

\medskip

Overall, sparsification of neural constitutive models has been widely adopted, and it has been shown that suitable regularization can reduce large FFNN-based constitutive models to compact ``single-line'' equations with only a few non-zero model parameters. In particular, component-wise $L_p$ regularization with $p\leq 1$ appears a reasonable choice for this. However, when the considered material behavior requires a larger amount of model parameters, it might be preferable to remain the matrix-vector representation of the neural constitutive model. In such cases, pruning the model equations means reducing the size of the weight matrices, i.e., deleting rows or columns that contain only zeros. Component-wise regularization does not directly promote this, since it penalizes the individual parameters, which can result in an irregular pattern of zeros of the parameter matrices. The number of non-zero parameters is then reduced, but the size of the matrix-vector operations is not, so evaluating the model still involves multiplications with a large number of zeros. This can be addressed by \emph{structured sparsification}, which is well-established in machine learning \parencite{bach_optimization_2011,kolb_smoothing_2026,hoefler2021}. Thereby, groups of parameters, e.g., all weights associated with one neuron, are penalized together, such that entire rows and columns can be removed from the FFNN parameter matrices. The result is a smaller network with dense weight matrices. To the best of our knowledge, structured sparsification has not yet been investigated for FFNN-based neural constitutive models.

\medskip

In this work, we investigate structured sparsification for hyperelastic PANN constitutive models. For this, we employ the PANN model proposed by \textcite{klein2022a,linden2023}. We consider both polyconvex and non-polyconvex models, which we apply to homogenization data of transversely isotropic rank-one laminates with different phase contrasts. Moreover, we employ one of the sparsification approaches proposed by \textcite{kolb_smoothing_2026}. For this, we assign gates to all inputs and neurons, and apply a $L_2$ penalty to both the gates and the weights, which is a smooth contribution to the loss function. Following \textcite{kolb_smoothing_2026}, this penalty drives entire groups of parameters towards zero. As a consequence, after training, entire rows and columns of the FFNN parameter matrices can be removed. The remaining structure of this manuscript is as follows. In~\cref{sec:basics}, we introduce the theoretical background of hyperelastic PANN modeling an the employed sparsification approach, which is followed by numerical examples in~\cref{sec:examples} and the conclusion in~\cref{sec:conc}

Regarding notation, single contractions between $\bA\in\bbR^{n_1\times n_2}$, $\bB\in\bbR^{n_2\times n_3}$, $\ba\in\bbR^{n_2}$, and $\bb\in\bbR^{n_2}$ are denoted by $(\bA\cdot\bB)_{ij}=A_{ik}B_{kj}\in\bbR^{n_1\times n_3}$, $(\bA\cdot\bb)_{i}=A_{ik}b_{k}\in\bbR^{n_1}$, and $\ba\cdot\bb=a_{k}b_{k}\in\bbR$, respectively, where Einstein's summation convention is applied and \(n_1,\:n_2,\:n_3\in\mathbb{N}^+\). Furthermore, $\norm{\bA}=\sqrt{A_{ij}A_{ij}}$ and $\norm{\ba}=\sqrt{a_ia_i}$ denote the Frobenius norm and the Euclidean norm, respectively.

\section{Hyperelastic PANN modeling}\label{sec:basics}

In this section, we introduce the basics of hyperelastic PANN modeling. The general model formulation is introduced in~\cref{sec:PANN}. The representation of the hyperelastic potential in terms of feed-forward neural networks (FFNNs) is discussed in~\cref{sec:neural_pot}, and the model training without $L_p$ regularization is introduced in~\cref{sec:training_without}.

\subsection{General model formulation}\label{sec:PANN}

We consider compressible hyperelastic constitutive models in three dimensions,\footnote{For a general introduction to hyperelasticity, to the construction of invariants, and to invariants in the context of polyconvexity, the reader is referred to \textcite{linden2023}, \textcite{riemer2026}, and \textcite{KLEIN2027106815}, respectively.} where the first Piola-Kirchhoff stress $\bP$ is given as the gradient of a hyperelastic potential $\psi$ with respect to the deformation gradient $\bF$, i.e.,
\begin{equation}\label{eq:stress}
    \bP=\partial_{\bF}\psi(\bcI(\bF))\,.
\end{equation}
Here, the potential is formulated in terms of a set of invariants $\bcI$ instead of $\bF$ directly. Throughout this work, we assume transversely isotropic material symmetry ($\cD_{\infty}$) and use the invariants
\begin{equation}\label{eq:invs_1}
    \bcI=\Big(\IoneISO,\,\ItwoISO,\,J,\,\IoneTI,\,\ItwoTI,\,-J,\,\IoneISO-\IoneTI,\,\ItwoISO-\ItwoTI\Big)\in\bbR^8\,,
\end{equation}
with
\begin{equation}\label{eq:invs_2}
    \begin{aligned}
        \IoneISO=\norm{\bF}^2\,,\quad \ItwoISO=\norm{\bH}^2\,,\quad \IoneTI=\norm{\bF\cdot\bG}^2\,,\quad \ItwoTI=\norm{\bH\cdot\bG}^2\,,
    \end{aligned}
\end{equation}
where $\bH=\cof\bF$ and $J=\det\bF$. Furthermore, $\bG=\bn_{0}\otimes\bn_{0}$ denotes a transversely isotropic structural tensor, where the unit vector $\bn_{0}$ describes the preferred direction of the material. Since the stress is defined as a gradient field in~\cref{eq:stress}, thermodynamic consistency in the form of energy conservation is fulfilled a priori, while the use of the invariants in~\cref{eq:invs_1} ensures material symmetry and objectivity. 

Moreover, we consider polyconvex potentials which can be expressed as
\begin{equation}
    \psi(\bcI(\bF))=\cP(\bF,\,\bH,\,J)\,,
\end{equation}
where $\cP$ is a convex function~\parencite{Ball1976,Ball1977}. Polyconvexity is linked to existence theorems in finite elasticity and can improve the numerical stability and robustness of simulations in which the constitutive model is applied \parencite{Schroeder_Neff_Balzani_2005}. All invariants in $\bcI$ are polyconvex, i.e., convex in $(\bF,\,\bH,\,J)$~\parencite{Schroeder2003}. Thus, a sufficient condition for polyconvexity of $\psi$ is that it is convex and component-wise monotonically increasing in $\bcI$~\parencite[Sec.~2.4]{KLEIN2027106815}. Note that the last three entries of $\bcI$ are not independent, as they are linear combinations of the first five. Including these redundant arguments is motivated by the formulation of polyconvex models, where such additional inputs can facilitate the model formulation or increase the flexibility of the model, as discussed in detail in~\textcite[Sec.~2.4]{KLEIN2027106815}. In the following, the potential $\psi$ is represented by an FFNN with the invariants $\bcI$ as inputs.

\subsection{Neural energy potential}\label{sec:neural_pot}

We consider scalar-valued FFNNs given as the mapping
\begin{equation}
    \cF\colon\bbR^{n_0}\rightarrow\bbR\,,\qquad \bx\mapsto\cF(\bx)\,,
\end{equation}
with the $n_0$-dimensional input $\bx$, and where the output $\cF$ is recursively defined as
\begin{equation}\label{eq:FFNN}
\begin{aligned}
{\mathrlap{\bx^{(0)}}\phantom{\bx^{(h)}}}&=\bx&&\in\bbR^{n_0}\,,
\\
{\mathrlap{\bx^{(h)}}\phantom{\bx^{(h)}}}&=\SP\big(\bW^{(h)}\cdot\bx^{(h-1)}+\bb^{(h)}\big)&&\in\bbR^{n_h}\,,\quad h=1,\,\dotsc,\,H \,,
\\
{\mathrlap{\cF}\phantom{\bx^{(h)}}}&=\bw^{(H+1)}\cdot\bx^{(H)}&&\in\bbR\,.
\end{aligned}
\end{equation}
Here, $H$ denotes the number of hidden layers, $n_h$ the number of neurons in each hidden layer, and $\SP(x)=\log(1+\exp(x))$ the softplus activation function, which is applied component-wise. The trainable parameters of the network are the weights $\bW^{(h)}\in\bbR^{n_h\times n_{h-1}}$, $\bw^{(H+1)}\in\bbR^{n_H}$ and the biases $\bb^{(h)}\in\bbR^{n_h}$, which are collected in $\bp^{\cW}$ and $\bp^{\cB}$,  respectively, with \(\boldsymbol{k}=(\bp^{\cW},\bp^{\cB})\). Since the softplus function is convex and monotonically increasing, $\cF$ is convex and monotonically increasing in each component of $\bx$ if all weights are non-negative, i.e., $k^{\cW}_i\geq 0$~\parencite{Amos2017}. 

Based on the FFNN in~\cref{eq:FFNN}, we can formulate a neural energy potential as
\begin{equation}\label{eq:psi_NN}
    \psi^{\text{NN}}(\bcI)=\cF(\bcI)\,,
\end{equation}
where the $n_0=8$ invariants in~\cref{eq:invs_1} serve as inputs of the FFNN. No bias is used in the output layer, since a constant in the potential does not affect the stress. If all weights of the FFNN in~\cref{eq:FFNN} are non-negative, the neural energy potential is polyconvex \parencite{klein2022a}. The resulting model fulfills thermodynamic consistency, objectivity, material symmetry, and polyconvexity a priori. We also consider unrestricted models, in which the weights are not constrained, and the fulfillment of polyconvexity is not guaranteed. 

To also fulfill the stress normalization condition $\bP\rvert_{\bF=\bI}=\bnull$, for both polyconvex and unrestricted models, the neural energy potential is complemented by the stress normalization term introduced in~\textcite[Sec.~3.3.2]{linden2023}, which does not introduce additional trainable parameters and is compatible with all aforementioned constitutive conditions. Furthermore, we add the polyconvex growth term $\alpha\,(J+J^{-1}-2)^2$ with $\alpha=1\,\text{MPa}$ to the neural energy potential, which ensures that the potential tends to infinity for $J\rightarrow 0^+$~\parencite{klein2022a}.

\subsection{Model training without $L_p$ regularization}\label{sec:training_without}

To calibrate the PANN model, we consider datasets of the form
\begin{equation}
    \cD=\Big\{\big({^1}\bF,\,{^1}\bP\big),\,\dotsc,\,\big({^m}\bF,\,{^m}\bP\big)
    \Big\}\,,
\end{equation}
which consist of $m$ tuples of deformation gradients and corresponding first Piola-Kirchhoff stresses. Each dataset is split into a calibration dataset with $m_{\text{c}}$ data points and a test dataset with $m_{\text{t}}=m-m_{\text{c}}$ data points. The model is fitted to the calibration dataset by minimizing the mean squared error (MSE)
\begin{equation}\label{eq:data_loss}
    \cL_{\text{data}}(\bp)=
    \frac{1}{9\,m_{\text{c}}\,\text{MPa}^2} \sum_{i=1}^{m_{\text{c}}} \big\| {{^i}\bP} - \bP^\text{model}({{^i}\bF};\,\bp) \big\|^2\,,
\end{equation}
where $\bP^\text{model}$ denotes the stress predicted by the PANN model. Training a FFNN on its gradients is referred to as \emph{Sobolev training} \parencite{vlassis2021,Czarnecki2017SobolevTF}. 

In a basic training scenario, the model is calibrated by minimizing only the data loss $\cL_{\text{data}}$, i.e., without an additional $L_p$ regularization term in the loss function. In~\cref{sec:reg}, the data loss is complemented by such a regularization term to induce structured sparsity.

\section{Structured sparsification of PANN models}\label{sec:sparsification}

In this section, we introduce the structured sparsification of PANN models employed in this work. FFNNs with gates are introduced in~\cref{sec:gates}, followed by the regularization as well as the pruning and second training round of the networks in~\cref{sec:reg,sec:pruning}.

\begin{figure}[t]
\centering
\includegraphics[width=0.9\columnwidth]{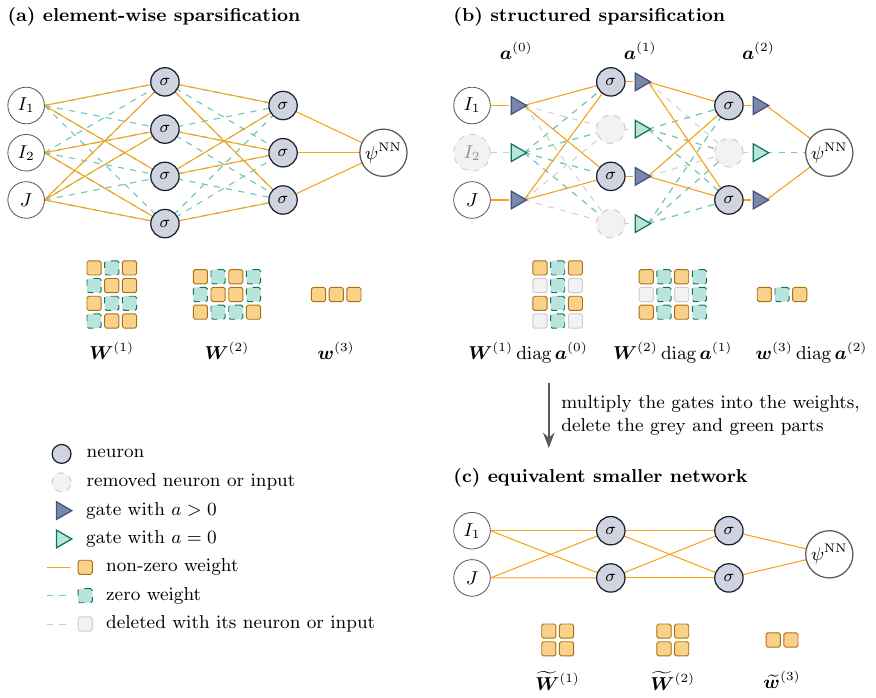}
 \caption{Element-wise vs. structured sparsification. (a) Individual zero weights leave the network size unchanged. (b) Closed gates zero entire columns of the weights, so the corresponding inputs and neurons can be removed. (c) Equivalent reduced network.}
\label{fig:key_idea}
\end{figure}

\subsection{FFNNs with gates}\label{sec:gates}

To realize structured sparsification, the FFNN in~\cref{eq:FFNN} is extended by gates, i.e.,
\begin{equation}\label{eq:FFNN_1}
\begin{aligned}
{\mathrlap{\bx^{(0)}}\phantom{\bx^{(h)}}}&=\operatorname{diag}\big(\ba^{(0)}\big)\cdot\bx&&\in\bbR^{n_0}\,,
\\
{\mathrlap{\bx^{(h)}}\phantom{\bx^{(h)}}}&=\operatorname{diag}\big(\ba^{(h)}\big)\cdot\SP\big(\bW^{(h)}\cdot\bx^{(h-1)}+\bb^{(h)}\big)&&\in\bbR^{n_h}\,,\quad h=1,\,\dotsc,\,H \,,
\\
{\mathrlap{\cF}\phantom{\bx^{(h)}}}&=\bw^{(H+1)}\cdot\bx^{(H)}&&\in\bbR\,,
\end{aligned}
\end{equation}
with the gates $\ba^{(h)}\in\bbR^{n_h}$. Each gate scales one input ($h=0$) or the output of one neuron ($h\geq 1$). The trainable parameters are now given by $\bp=\big(\bp^{\cW},\,\bp^{\cB},\,\bp^{\cA}\big)$, where all gates are collected in $\bp^{\cA}$. The FFNN with gates in~\cref{eq:FFNN_1} can equivalently be expressed as the standard FFNN in~\cref{eq:FFNN} with the effective weights
\begin{equation}\label{eq:FFNN_2}
\widetilde{\bW}^{(h)}=\bW^{(h)}\cdot\operatorname{diag}\big(\ba^{(h-1)}\big)\,,\quad h=1,\,\dotsc,\,H\,,\qquad \widetilde{\bw}^{(H+1)}=\bw^{(H+1)}\cdot\operatorname{diag}\big(\ba^{(H)}\big)\,.
\end{equation}
Thus, $\cF$ is convex and monotonically increasing in each component of $\bx$ if all weights and gates are non-negative, i.e., $k^{\cW}_i,\,k^{\cA}_i\geq 0$, and the construction of polyconvex PANN models described in~\cref{sec:basics} carries over to FFNNs with gates. Moreover, for $a_i^{(h)}=1$, the standard FFNN in~\cref{eq:FFNN} is recovered. 

\subsection{Structured regularization}\label{sec:reg}

Following~\textcite{kolb_smoothing_2026}, the parameters of the FFNN with gates are optimized by minimizing the loss function
\begin{equation}\label{eq:loss}
    \cL(\bp)=\cL_{\text{data}}(\bp)+\lambda\,\cL_{\text{reg}}(\bp)\,,
\end{equation}
where $\cL_{\text{data}}$ is the data loss defined in~\cref{eq:data_loss}, $\cL_{\text{reg}}$ is a regularization term, and $\lambda\geq 0$ is a penalty parameter. We choose
\begin{equation}\label{eq:reg}
    \cL_{\text{reg}}(\bp)= \big\|\bp^{\cW}\big\|^2 + \big\|\bp^{\cA}\big\|^2\,,
\end{equation}
which corresponds to an $L_2$ regularization of the weights and gates. The biases are not regularized, as the removal of an input or neuron only requires its gate or its outgoing weights to vanish, cf.~\cref{sec:pruning}. Note that penalizing only the gates would not be that effective, as very large weights could compensate for very small gates.

At first glance, it might be surprising that sparsity can be induced by an $L_2$ regularization, as the literature on sparsification of neural constitutive models reports sparsity for $p\leq 1$, but not for $p=2$, cf.~\cref{sec:intro}. However, these works employ component-wise regularization, i.e., each parameter is penalized individually, e.g., by
    \begin{equation}
{\cL}^*_{\text{reg}} =\sum_i \lvert k_i^{\cW} \rvert^p\,,
\end{equation}
for which $p\leq 1$ is a reasonable choice to induce sparsity. The situation is different for structured regularization, since the gates couple the parameters multiplicatively, cf.~\cref{eq:FFNN_2}. As shown by~\textcite{kolb_smoothing_2026}, the regularization in~\cref{eq:reg} is therefore closely linked to penalizing the norms of the columns of the weight matrices of a FFNN without gates, i.e.,\footnote{This is also referred to as group lasso~\parencite{yuan_model_2006}.}
\begin{equation}
{\cL}^{**}_{\text{reg}} =\sum_j \lVert {\bw}^*_j \rVert\,,
\end{equation}
where ${\bw}^*_j$ denotes the columns of ${\bW}^{(h)}$ as well as the components of ${\bw}^{(H+1)}$ of~\cref{eq:FFNN}. We provide a schematic comparison of component-wise and structured regularization in~\cref{fig:key_idea}.

\subsection{Pruning and second training round}\label{sec:pruning}

Due to~\cref{eq:FFNN_2}, a vanishing gate $a_i^{(h)}=0$ results in a zero $i$-th column of $\widetilde{\bW}^{(h+1)}$, or a zero $i$-th component of $\widetilde{\bw}^{(H+1)}$ for $h=H$.\footnote{Since the weights are regularized as well, zero columns may also appear directly in $\bW^{(h)}$ and $\bw^{(H+1)}$.} Such zero columns can be deleted without altering the predictions of the FFNN. Moreover, the corresponding entry of $\bx^{(h)}$, which is only multiplied by zeros, can be deleted as well. For $h=0$, this removes an input, and for $h\geq 1$, it removes a neuron of the $h$-th hidden layer together with the corresponding row of $\widetilde{\bW}^{(h)}$ and entry of $\bb^{(h)}$. A schematic visualization of this procedure is given in~\cref{fig:key_idea}b,c.

For the numerical investigations of this work, non-negativity constraints on model parameters are realized by clipping after each weight update, i.e., each non-negative parameter is represented as $w^+=\max(w,\,0)\in\bbR_{\geq 0}$ with an unconstrained variable $w\in\bbR$. In contrast to alternative approaches such as $w^+=\exp(w)$, which never yield values of exactly zero, this allows model parameters to become exactly zero. In such cases, one could define a tolerance below which the weights are pruned. Although the clipping is not differentiable at $w=0$, it is commonly applied in practice and proved to be robust throughout our numerical investigations.

Besides removing inputs and neurons, the regularization also penalizes the magnitude of the remaining weights, which might reduce the accuracy of the model. Therefore, after merging the gates into the effective weights, the reduced network is subsequently retrained without regularization, i.e., by minimizing~\cref{eq:loss} with $\lambda=0$ and fixed gates, while its structure is kept fixed. 

Finally, redundant inputs are removed. Since the first hidden layer before the activation function is linear in its inputs and the last three invariants in~\cref{eq:invs_1} are linear combinations of the first five, the corresponding columns of the weight matrix in the first hidden layer can be merged. For instance, the columns corresponding to $J$ and $-J$ can be combined into a single column for $J$. In this way, further inputs are removed without altering the predictions of the FFNN, such that at most five inputs remain. In the polyconvex case, merged weights can then become negative. However, as the function represented by the network does not change, the reduced model remains polyconvex.

With the numbers of remaining inputs $n_0$ and neurons $n_h$, its number of trainable parameters is given by
\begin{equation}\label{eq:n_params}
    n_{\text{p}}=\sum_{h=1}^{H}n_h\,\big(n_{h-1}+1\big)+n_H\,,
\end{equation}
which comprises the weights and biases of the hidden layers as well as the weights of the output layer.

\section{Numerical examples}\label{sec:examples}

In this section, we present numerical examples on the proposed structured sparsification of hyperelastic PANN models. The considered data, the models, and the calibration and implementation details are described in~\cref{sec:data_model}.


\subsection{Data and model preparation}\label{sec:data_model}

\begin{figure}[t]
        \centering
         \includegraphics[width=3.6cm]{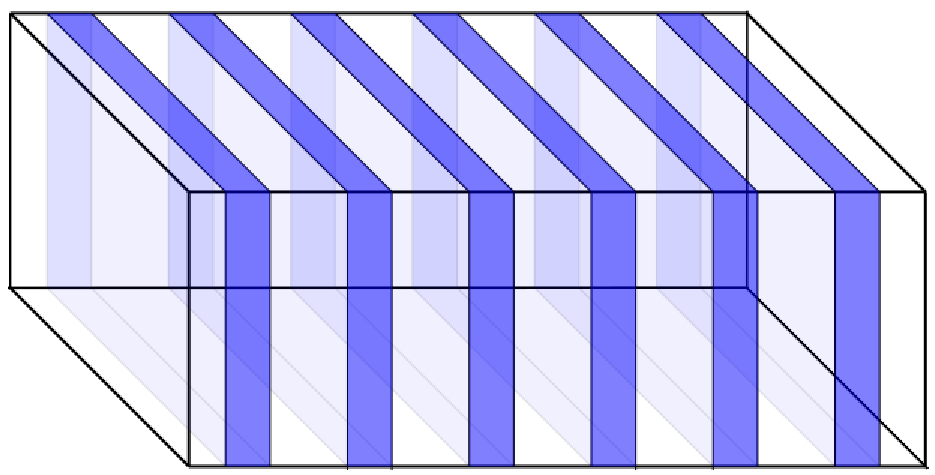}
         \caption{Rank-one laminate microstructure (ROL). Image from \textcite{klein2024a}.}
	\label{fig:ROL}
\end{figure}

We now provide information on the considered data and model preparation.

\paragraph{Data} We consider data from~\textcite{klein2026b}, which were generated by the semi-analytical homogenization of a biphasic rank-one laminate (ROL) following the homogenization method proposed by~\textcite{Debotton_2005}. The ROL consists of alternating layers of two phases, which results in a transversely isotropic effective material behavior ($\cD_{\infty}$) with the preferred direction $\bn_{0}$ normal to the layers, cf.~\cref{fig:ROL}. Both phases are described by the isotropic Mooney-Rivlin model
\begin{equation}\label{eq:micro_behavior_ROL}
    \psi_{\circ}=\frac{1}{2}\Big[\mu_{1,\circ}\IoneISO+\mu_{2,\circ}\ItwoISO-2(\mu_{1,\circ}+2\mu_{2,\circ})\log J+\kappa_{\circ}(J-1)^2\Big]\,,
\end{equation}
where the index $\circ\in\{a,\,b\}$ assigns the material model to one of the phases, and both phases have equal volume fractions. For phase~$a$, the material parameters are $\mu_{1,a}=\mu_{2,a}=0.5\,\text{MPa}$ and $\kappa_a=50\,\mu_{1,a}$. For phase~$b$, they are scaled by the phase contrast $f_m$, i.e., $\mu_{1,b}=\mu_{2,b}=f_m\,\mu_{1,a}$ and $\kappa_b=f_m\,\kappa_a$. For the phase contrast $f_m$, a low value of 1.5 and a high value of 10 are employed. 

The deformation gradients were sampled as provided by~\textcite{klein2026b} with the concentric sampling strategy proposed by~\textcite{kunc2019}, which covers a wide range of physically admissible deformation modes, see also~\textcite[App.~B]{klein2022b} for a brief introduction to the method. The datasets consist of $m=11\,000$ data points for the low and $m=8\,236$ data points for the high phase contrast. The smaller number for the high phase contrast results from excluding all data points at which the homogenized material is not elliptic, following~\textcite[Sec.~4.1.1]{klein2026b}. While \textcite{klein2026b} also provide data for the tangent $\bbA=\partial_{\bF}\bP$, only the stress $\bP$ is considered in this work. For each dataset, $m_{\text{c}}=800$ randomly selected datapoints are used for calibration, and the remaining $m_{\text{t}}=m-m_{\text{c}}$ datapoints are used for testing.

\paragraph{Models} We consider transversely isotropic hyperelastic PANN models as introduced in~\cref{sec:basics}. Polyconvex models are employed for both phase contrasts, where both the weights and gates are non-negative. For the high phase contrast, unrestricted models are considered in addition, in which the weights are not restricted in sign, while the gates are non-negative. For comparability, all models use the invariants in~\cref{eq:invs_1} as inputs, although the first five invariants would be sufficient for the unrestricted models, as the remaining ones are linear combinations of them. Models trained without $L_p$ regularization employ the FFNN in~\cref{eq:FFNN}, and models trained with structured $L_p$ regularization employ the FFNN with gates in~\cref{eq:FFNN_1}. The pruning and retraining described in~\cref{sec:pruning} are applied to all models, including those trained without regularization, where zero columns can arise from the clipping. 

\paragraph{Initial architectures and penalty parameters} Networks with one and two hidden layers are considered, where both hidden layers of the two-layer networks have the same initial number of neurons. Without regularization, i.e., for $\lambda=0$, networks with $n_h\in\{1,\,2,\,4,\,\dotsc,\,256\}$ nodes and $n_h\in\{1,\,2,\,4,\,\dotsc,\,64\}$ nodes are used for single-layered and two-layered networks, respectively. They serve as a reference that represents the classical hyperparameter study, in which the network size is varied directly, cf.~\cref{sec:intro}. With regularization, we start from networks with 16, 64, and 256 neurons for one hidden layer and with 16, 32, and 64 neurons per hidden layer for two hidden layers. The largest initial networks thus have a comparable number of parameters, i.e., roughly $2\,500$ and $4\,800$, respectively. Smaller initial networks are not considered, as sparsification is intended for networks whose size is chosen generously, while small network sizes are already covered by the networks without regularization. The penalty parameter is varied over four orders of magnitude, $\lambda\in\{10^{-6},\,10^{-5},\,10^{-4},\,10^{-3}\}$, ranging from a weak to a strong regularization, so that the trade-off between accuracy and model size is covered.

\paragraph{Calibration and implementation details} The models are implemented in KLAX~0.1.3 (\url{https://drenderer.github.io/klax/}). The weights are initialized with He normal initialization, the biases with zeros, and the gates with ones. Models with $L_p$ regularization are trained, pruned, and retrained as described in~\cref{sec:pruning}. Due to the clipping strategy described in~\cref{sec:neural_pot}, gates and weights can become exactly zero. Thus, only inputs and neurons whose columns of the effective weights are exactly zero are removed, such that the pruning does not alter the model predictions. Models without $L_p$ regularization are calibrated with the same two-stage procedure but without gates. For the polyconvex models without $L_p$ regularization, zero columns arising from the clipping of the weights are pruned after the first stage. In both stages, the Adam optimizer is used with a batch size of 32. The first and second stages consist of $2\cdot 10^{5}$ and $10^{5}$ training steps, respectively, with initial learning rates of $2\cdot 10^{-3}$ and $10^{-3}$, which are multiplied by $0.1$ after $90\,\%$ and $95\,\%$ of the training steps, respectively. For each combination of network size and penalty parameter, five models are trained. Each of them is calibrated on a different random selection of data points, and the same five splits into calibration and test data are used for all network sizes and penalty parameters to ensure consistent comparisons.

\begin{figure}[t]
        \centering
         \includegraphics[width=0.95\columnwidth]{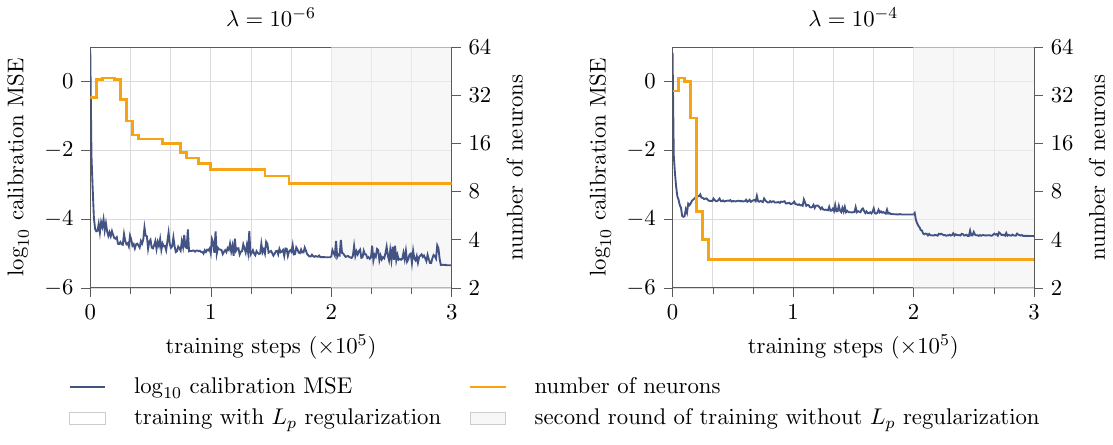}
         \caption{Representative training histories for polyconvex PANNs with one hidden layer applied to the ROL with low phase contrast, with $\lambda=10^{-6}$ and $\lambda=10^{-4}$, respectively.}
	\label{fig:training_histories}
\end{figure}

\begin{figure}[t]
        \centering
         \includegraphics[width=0.86\columnwidth]{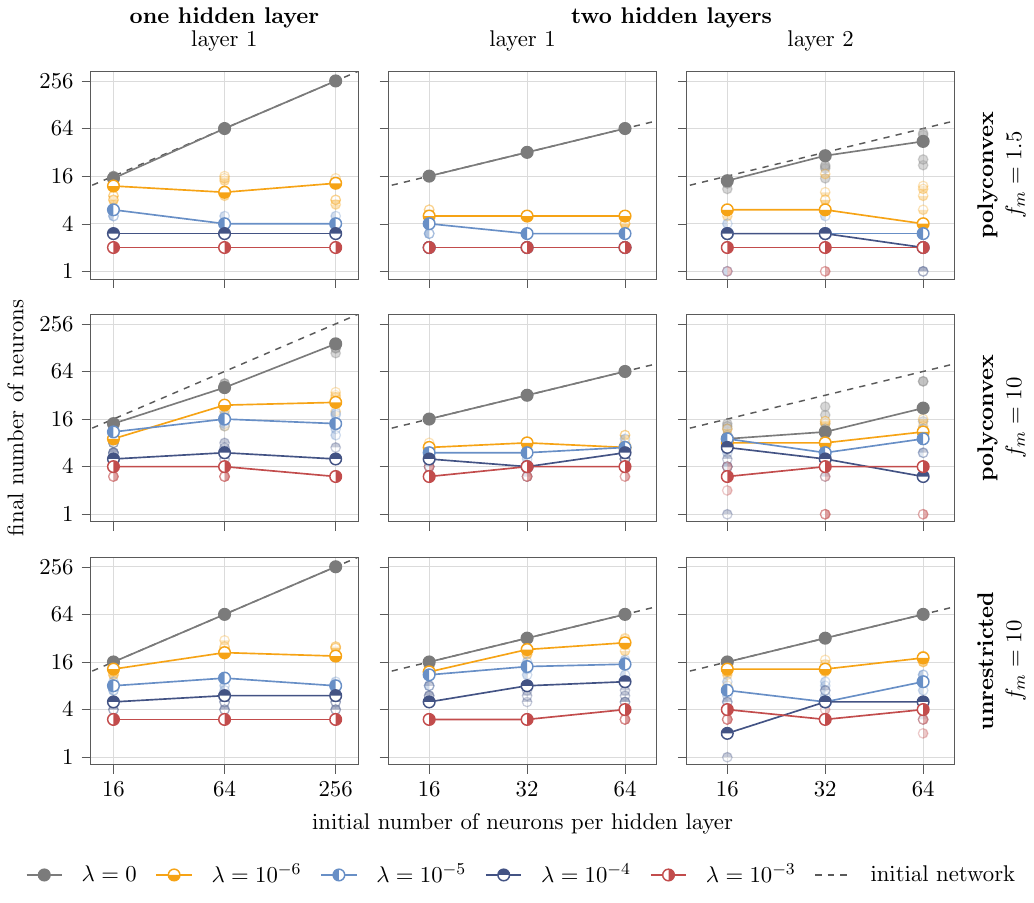}
         \caption{Final network sizes. The connected, opaque markers denote the respective model instances with the lowest calibration MSE, while the remaining model instances are visualized with reduced opacity.
         }
	\label{fig:neurons}
\end{figure}

 
\subsection{Training behavior}\label{sec:training_behavior}

\Cref{fig:training_histories} shows two representative training trajectories for the penalty parameters $\lambda=10^{-6}$ and $\lambda=10^{-4}$, both for a polyconvex PANN with one hidden layer of initially 64 neurons, calibrated to the ROL data with low phase contrast. Notably, in both cases the number of neurons at the beginning of the training is considerably below 64. This is likely caused by output-layer weights that are already zero at initialization, because of the pruning of weights initialized to negative values. Some of these neurons become active again during the first training steps, so the number of neurons temporarily increases before the sparsification influence sets in.

For $\lambda=10^{-6}$, the calibration MSE drops by about five orders of magnitude in the first training round, as the training mainly focuses on representing the underlying material behavior. The sparsification proceeds slowly and ends at a comparatively large number of neurons, as will be discussed in detail in~\cref{sec:size}. The second round of training without $L_p$ regularization improves the calibration MSE only slightly. For $\lambda=10^{-4}$, the network size is considerably reduced at the beginning of the training process, which leads to a temporary increase in the data loss. Here, the second round of training without $L_p$ regularization considerably improves the model performance in predicting the underlying material behavior. Overall, this suggests that a larger penalty parameter can be useful to reduce the network size, while a subsequent second round of training can ensure a good prediction accuracy of the model if enough parameters to describe the material behavior remain.

 
\subsection{Final network size}\label{sec:size}
 
\Cref{fig:neurons} shows the final number of neurons per hidden layer over the initial number of neurons. For instance, for the polyconvex model with one hidden layer and $f_m=1.5$, the networks starting from 16, 64, and 256 neurons are reduced to about 10 to 12 neurons for $\lambda=10^{-6}$, to 4 to 6 neurons for $\lambda=10^{-5}$, and to 3 and 2 neurons for $\lambda=10^{-4}$ and $\lambda=10^{-3}$, respectively, regardless of the initial size. Thus, a network with 256 initial neurons is reduced to the same size as a network with 16 initial neurons, and the larger the penalty parameter, the smaller the final network. For two hidden layers, both hidden layers are reduced to similar sizes, e.g., to 2 to 6 neurons per layer for the polyconvex model with $f_m=1.5$. Only for the smallest penalty parameter $\lambda=10^{-6}$, the final size partly depends on the initial size. For instance, for the polyconvex model with one hidden layer and $f_m=10$, about 9 neurons remain when starting from 16 neurons, compared to about 23 to 25 neurons when starting from 64 or 256 neurons. Similarly, for the unrestricted model with one hidden layer, the network starting from 16 neurons keeps about 13 of them, compared to about 18 to 20 neurons for the larger initial networks. Moreover, for the same penalty parameter, more neurons remain for the high phase contrast than for the low phase contrast. For instance, for one hidden layer and $\lambda=10^{-5}$, about 11 to 16 neurons remain for the polyconvex model with $f_m=10$, compared to 4 to 6 neurons for $f_m=1.5$, cf.~\cref{sec:accuracy}. Consequently, the choice of the network size, which otherwise requires expert knowledge or a hyperparameter study, cf.~\cref{sec:intro}, is replaced by the choice of the penalty parameter, while the initial network only has to be chosen sufficiently large. Thus, with structured regularization and sparsification, the final network size is largely independent of the initial network size and mainly controlled by the penalty parameter.

The pruning of the inputs is largely independent of whether one or two hidden layers are considered, but differs between the polyconvex and unrestricted models as well as between the low and high phase contrast of the ROL. For the polyconvex PANNs calibrated to the ROL with low phase contrast, roughly $40\%$ of the trained models retain four inputs, while the others retain all five inputs. For the polyconvex PANNs calibrated to the ROL with high phase contrast, roughly $75\%$ of the trained models retain four inputs, while the others retain all five inputs. For the unrestricted PANNs, almost all trained models retain all five inputs. Notably, the pruned invariants usually still enter the model through one of the redundant inputs, e.g., if the input $\IoneISO$ is pruned, but the input $\IoneISO-\IoneTI$ is retained. 

Without regularization, the unrestricted networks keep all of their neurons, whereas the polyconvex networks lose some of their neurons, in particular in the last hidden layer. For instance, for two hidden layers and 64 initial neurons per layer, only about 45 and 20 neurons remain in the second hidden layer for $f_m=1.5$ and $f_m=10$, respectively, while the first hidden layer keeps all of its neurons. This is caused by the clipping of the non-negative weights of the polyconvex models. A neuron of the last hidden layer is removed as soon as its single output weight vanishes, whereas a neuron of the first hidden layer of a two-layer network is only removed if all of its outgoing weights vanish, which explains why mainly the last hidden layer is affected. 

\begin{figure}[t]
        \centering
         \includegraphics[width=0.7\columnwidth]{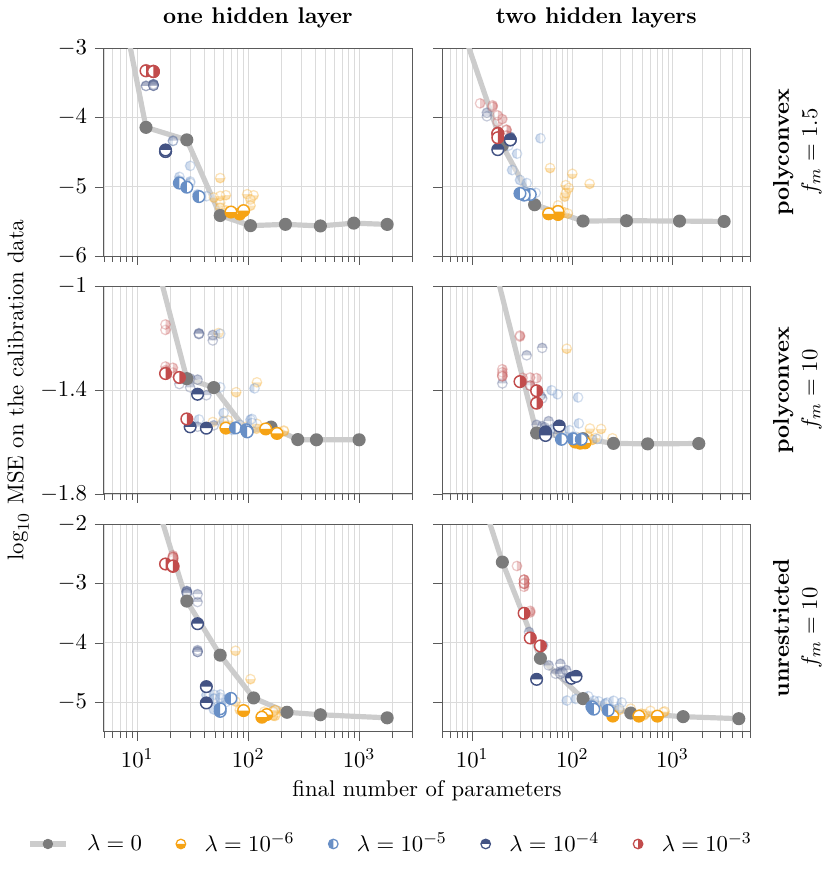}
         \caption{Final $\log_{10}$ calibration MSE depending on the final number of NN parameters. For each combination of PANN model type, considered data, and initial network size, the model instance with the lowest calibration MSE is shown by opaque markers, while the remaining instances are shown with reduced opacity.         
         }
	\label{fig:MSE}
\end{figure}

 
\subsection{Accuracy and network size}\label{sec:accuracy}
 
\Cref{fig:MSE} shows the calibration MSE over the final number of parameters, cf.~\cref{eq:n_params}. Note that in all investigated cases, the values of the test MSE are close to the values of the calibration MSE, i.e., no overfitting is observed. For the networks without regularization, the MSE decreases with increasing network size until it reaches a plateau of about $10^{-5.5}$ for the polyconvex models with $f_m=1.5$, $10^{-1.6}$ for the polyconvex models with $f_m=10$, and $10^{-5.2}$ for the unrestricted models with $f_m=10$. For the polyconvex models with $f_m=10$, the plateau is more than three orders of magnitude above the one of the unrestricted models, which is because the employed polyconvex PANN model can be overly restrictive when applied to the ROL data with high phase contrast, as discussed in~\textcite{klein2026b}. The plateau is reached with about 100 to 400 parameters. Thus, the largest networks with about $1\,800$ and $4\,800$ parameters for one and two hidden layers, respectively, are clearly overparametrized for the considered data. Overall, for a given number of parameters, the sparsified networks are at least as accurate as the networks trained without regularization, and for one hidden layer, they are up to almost one order of magnitude more accurate. 
 
The sparsified networks reflect the trade-off between network size and accuracy, which is controlled by the penalty parameter. For $\lambda=10^{-6}$, the networks are reduced to about 55 to 700 parameters and reach MSE values at or close to the plateau. With increasing penalty parameter, the networks become smaller, while the MSE increases. For $\lambda=10^{-3}$, networks with about 12 to 50 parameters remain, whose MSE is up to two and a half orders of magnitude above the plateau. Thus, a too small penalty parameter results in accurate networks that are larger than required, while a too large penalty parameter results in very small networks with a considerably reduced accuracy.
 
Most importantly, the sparsified networks lie on or below the grey curve of the networks without regularization in most cases. The improvement is most pronounced for one hidden layer and small to intermediate network sizes. For instance, for the polyconvex model with $f_m=1.5$, the sparsified networks with about 25 to 35 parameters ($\lambda=10^{-5}$) reach MSE values of about $10^{-5.0}$ to $10^{-5.15}$, whereas the network without regularization with 28 parameters only reaches $10^{-4.3}$. Thus, starting from a large network and reducing it during training can result in better small networks than training small networks directly. For two hidden layers, the sparsified networks lie approximately on the curve of the networks without regularization, i.e., they are as accurate as networks of the same size trained without regularization. In this case, the main benefit of the sparsification is that the network size does not have to be chosen beforehand. An exception is the strongest regularization $\lambda=10^{-3}$ for the polyconvex model with one hidden layer and $f_m=1.5$. Here, the sparsified networks consist of two neurons and about 14 parameters and reach an MSE of about $10^{-3.35}$, whereas the network without regularization with two neurons and 12 parameters reaches $10^{-4.15}$. This indicates that a too strong regularization can remove inputs or neurons that would be beneficial for the accuracy and cannot be recovered by the retraining. 

Note that the penalty parameter weighs the regularization term against the data loss, whose magnitude depends on the magnitude of the stresses and on how well the model class can represent the data. Thus, the same penalty parameter can result in different network sizes for different datasets and model classes. However, the stress magnitudes for the ROL with low and high phase contrast are in the same order of magnitude, and this effect is not further investigated here. A normalization of the data loss, e.g., by the variance of the stress data, could make the penalty parameter more transferable between different datasets. More generally, the larger number of remaining neurons for the high phase contrast suggests that more complex material behavior requires larger networks, in which case structured sparsification, which reduces the size of the weight matrices, would be particularly beneficial for the computational efficiency.

\begin{figure}[t]
        \centering
\includegraphics[width=0.85\columnwidth]{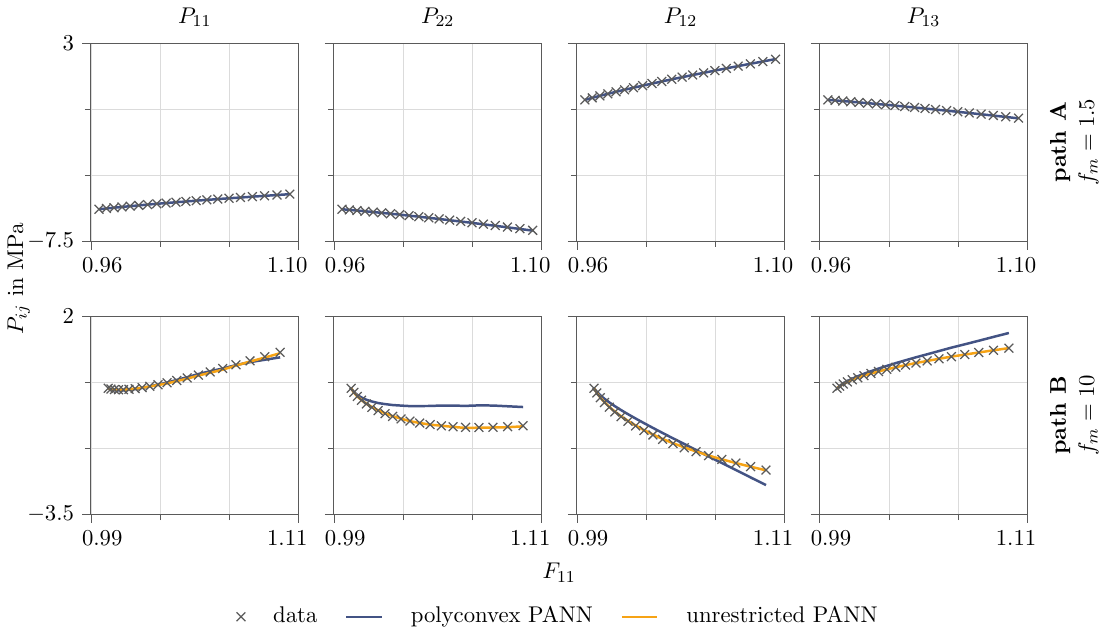}
         \caption{Representative stress predictions for two load paths for the models discussed in~\cref{sec:required_accuracy}.}
	\label{fig:stress}
\end{figure}

 
\subsection{Required accuracy}\label{sec:required_accuracy}

So far, the models were assessed by their calibration MSE, where the lowest values were reached by the networks without regularization and by the sparsified networks with small penalty parameters. However, the lowest achievable data loss is not required for the application of a constitutive model. For instance, in~\textcite[Sec.~4.3.2]{klein2026b}, a polyconvex PANN calibrated to the same ROL data with low phase contrast reached a $\log_{10}$MSE of about $-4$, and nonlinear FE simulations with this model virtually coincided with simulations based on the ground truth ROL behavior. Thus, intermediate MSE values in~\cref{fig:MSE} can already be sufficient for an accurate representation of the material behavior, such that a moderate penalty parameter can be chosen. To illustrate this, we consider three exemplary sparsified models, for which we provide the neural energy potential in the following. First, a polyconvex PANN with a single hidden layer\footnote{Recall that the negative weights for the polyconvex models are caused by combining the weights of redundant invariants after training, cf.~\cref{sec:pruning}. Note that the NN energy potentials are complemented by the stress normalisation and growth terms introduced in~\cref{sec:PANN}.}
\renewcommand{\arraystretch}{1.4}
\begin{equation}\label{eq:example_1}
   \psi^{\text{NN}}=\begin{bmatrix}
0.7292 & 3.7579 & 3.4314
   \end{bmatrix}
   \cdot
   \SP\left(\begin{bmatrix}
0.4174 & 0 & 0.4407 & 0 \\
0 & -7.0986 & 0.0008 & 0.0016 \\
0 & -7.4141 & 0.0023 & 0.0085 \\
    \end{bmatrix}
    \cdot
\begin{bmatrix}
   \ItwoISO \\
    J \\
    \IoneISO - \IoneTI \\
    \ItwoISO - \ItwoTI
\end{bmatrix}
+
\begin{bmatrix}
    1.0594 \\
    8.4032 \\
    6.0107 
\end{bmatrix}
    \right)\,,
\end{equation}
which we calibrated to the ROL with low phase contrast, where it achieved a $\log_{10}$MSE of $-4.47$ on the calibration data. Next, an unrestricted PANN with a single hidden layer
\begin{equation}
\begin{split}
   \psi^{\text{NN}}= & \begin{bmatrix}
1.4318 & 0.7068 & 7.2311 & 2.0904 & 8.3542
   \end{bmatrix}\\
   &\cdot
   \SP\left(\begin{bmatrix}
1.0916 & 2.1649 & -1.3060 & -0.5695 & 0.8753\\
-0.7051 & 0.6719 &  0.6789 & 2.6790 & 0.0033 \\
0.0619 & 5.9986 & 0.0072 & -0.0819 & -0.0709 \\
0.8661 & -0.2457 & -0.9661 & -5.1931 & 0.0920 \\
0.0575 & -5.1086 & 0.0094 & -0.0514 & -0.0702
    \end{bmatrix}
    \cdot
\begin{bmatrix}
    \IoneISO\\
    J \\
    \IoneTI \\
    \ItwoTI \\
    \ItwoISO - \ItwoTI
\end{bmatrix}
+
\begin{bmatrix}
    7.6320 \\
    -2.6615 \\
    -5.0675 \\
    1.2320 \\
    6.2187
\end{bmatrix}
    \right)
\end{split}
\end{equation}
which we calibrated to the ROL with high phase contrast, where it achieved a $\log_{10}$MSE of $-3.68$ on the calibration data. Finally, a polyconvex PANN with two hidden layers
\begin{equation}\label{eq:example_3}
\begin{split}
   \psi^{\text{NN}}= 
   &\begin{bmatrix}
0.1723 & 2.5432 & 0.7957 & 1.3273
   \end{bmatrix}
   \cdot
   \SP\left(\begin{bmatrix}
 0.6509 & 0 & 2.7269 & 0  \\
0.0008 & 2.7376 & 0 & 1.9794  \\
0 & 3.2123 & 0 & 0  \\
0 & 2.4667 & 0 & 0.3468  \\
    \end{bmatrix} 
    \right. \\
 &\left. \cdot 
 \SP \left(
 \begin{bmatrix}
-0.4547 & 6.4916 & 0 & 0 \\
5.738 & 0 & 0 & 0.0009 \\
-0.2657 & 0 & 3.1921 & 0 \\
-3.0981 & 0 & 0 & 0.3868 \\
    \end{bmatrix} 
    \cdot
    \begin{bmatrix}
    J\\
    \ItwoTI \\
    \IoneISO - \IoneTI\\
    \ItwoISO - \ItwoTI
    \end{bmatrix}
    +
    \begin{bmatrix}
    -9.1891 \\
    -5.1099 \\
    -3.7469\\
    2.379
\end{bmatrix}
 \right)
+
\begin{bmatrix}
    -5.7739 \\
    8.0389 \\
    -0.6233 \\
    -5.0346
\end{bmatrix}
    \right)
\end{split}
\end{equation}
which we calibrated to the ROL with high phase contrast, where it achieved a $\log_{10}$MSE of $-1.53$ on the calibration data.

Exemplary stress predictions for these models are visualised in~\cref{fig:stress}. For the polyconvex PANN applied to the ROL with low phase contrast and the unrestricted PANN applied to the high phase contrast, the stress prediction quality is excellent, and virtually no difference between data and model predictions can be observed.

Notably, for both cases, other models in~\cref{fig:MSE} reach considerably lower data losses, as the plateaus of the networks without regularization lie about one order of magnitude lower. In light of the excellent predictions of the models with intermediate data loss, this strongly suggests that the lowest achievable data loss is not required for an accurate representation of the material behavior. In contrast, the polyconvex PANN applied to the ROL with high phase contrast shows visible deviations. However, this is not caused by the sparsification, but reflects the practical limitations of the employed polyconvex model formulation for the ROL with high phase contrast~\parencite{klein2026b}.
 
\begin{remark}
As discussed in~\cref{sec:intro}, sparsified neural constitutive models can often be written as compact closed-form expressions without matrix-vector notation. In this spirit, \cref{eq:example_1} can equivalently be expressed as
    \begin{equation}
\begin{split}
    \psi^{\text{NN}}
    & = 0.7292 \cdot \SP \left(0.4174\ItwoISO + 0.4407(\IoneISO - \IoneTI) + 1.0594\right) \\
    & \quad + 3.7579 \cdot \SP \left(-7.0986 J + 0.0008(\IoneISO - \IoneTI) + 0.0016(\ItwoISO - \ItwoTI) + 8.4032\right) \\
    & \quad + 3.4314 \cdot \SP \left(-7.4141 J + 0.0023(\IoneISO - \IoneTI) + 0.0085(\ItwoISO - \ItwoTI) + 6.0107\right)\,.
\end{split}
\end{equation}
In principle, NN potentials with multiple hidden layers and more neurons, such as~\cref{eq:example_3}, could be written down in the same way, albeit with considerably more lines. Which representation is preferable thus depends on the purpose, e.g., the explicit form for interpreting small models and the matrix-vector notation for implementation.
\end{remark}

\section{Conclusion}\label{sec:conc}

In this work, we applied structured sparsification to hyperelastic PANN models. For this purpose, gates are assigned to all inputs and neurons of the FFNN, and an $L_2$ regularization is applied to the weights and gates. This regularization is closely linked to penalizing the norms of the columns of the effective weight matrices~\parencite{kolb_smoothing_2026}, and thus drives entire inputs and neurons to zero. As the non-negativity constraints are realized by clipping, gates and weights can become exactly zero, such that the corresponding inputs and neurons can be removed without a threshold and without altering the model predictions. After retraining, the result is a smaller, dense FFNN that can be evaluated with standard implementations. The approach was investigated for polyconvex and unrestricted PANN models with one and two hidden layers, calibrated to data of a rank-one laminate with low and high phase contrast. The main findings are:
\begin{itemize}
\setlength{\itemsep}{0pt}
    \item With structured regularization and sparsification, the final network size is largely independent of the initial network size and mainly controlled by the regularization penalty parameter.
    \item For a given number of parameters, the sparsified networks are at least as accurate as networks trained without regularization. 
    \item As to be expected, the penalty parameter controls the trade-off between network size and accuracy.
    \item The non-negativity constraints of the polyconvex models result in a partial sparsification even without regularization.
\end{itemize}
Thus, structured sparsification provides compact networks with small, dense weight matrices without an extensive hyperparameter study over the network size. The choice of the penalty parameter, which weighs the regularization term against the data loss, remains. However, penalty parameters between $10^{-6}$ and $10^{-4}$ all resulted in compact networks whose accuracy is within about one order of magnitude of the best accuracy achieved, such that this choice is not overly critical.

Finally, we emphasize that structured sparsification is one promising approach to downsizing neural constitutive models, but certainly not the only one. As discussed in~\cref{sec:intro}, component-wise regularization has also been shown to result in compact model equations. A main difference is that component-wise approaches require an $L_p$ regularization with $p\leq 1$ to induce sparsity, which is not differentiable at zero, non-convex for $p<1$, and even discontinuous for $p=0$. In contrast, the structured approach employs a smooth $L_2$ regularization, as the multiplicative gates turn it into a group-wise $L_1$ regularization of the effective weights, cf.~\cref{sec:reg}. For models as small as those obtained here, it is presumably of minor importance for the computational efficiency whether the model equations are written out explicitly or evaluated as small matrix-vector operations. 


\vspace*{3ex}

\noindent
\textbf{CRediT authorship contribution statement.}
\textbf{A.\ Kadereit:} Conceptualization, Methodology, Software, Formal analysis, Investigation, Visualization, Writing -- original draft, Writing -- review \& editing.
\textbf{D.K.\ Klein:} Conceptualization, Methodology, Software, Formal analysis, Investigation, Visualization, Supervision, Writing -- original draft, Writing -- review \& editing.
\textbf{M.\ Franke:} Supervision, Writing -- review \& editing.
\textbf{O.\ Weeger:} Conceptualization, Funding acquisition, Writing -- review \& editing.
\vspace*{1ex}

\noindent
\textbf{Conflict of interest.} The authors declare that they have no conflict of interest.
\vspace*{1ex}

\noindent
\textbf{Acknowledgment.} 
D.K.\ Klein and O.\ Weeger acknowledge the financial support provided by the Deutsche Forschungsgemeinschaft (DFG, German Research Foundation, project number 492770117) and the Graduate School of Computational Engineering at TU Darmstadt. M.\ Franke acknowledges the financial support provided by the Deutsche Forschungsgemeinschaft (DFG, German Research Foundation, project number 443238377).
\vspace*{1ex}

\noindent
\textbf{Data availability.}
The authors have no data to share.
\vspace*{1ex}

\noindent
\textbf{AI declaration.} During the preparation of this work, the authors used Claude (Anthropic) and ChatGPT (OpenAI) to support language editing, improve clarity and readability, and assist with code development and debugging. All scientific content, methodological choices, numerical results, and conclusions were developed, verified, and approved by the authors. The authors take full responsibility for the content of the manuscript.


\renewcommand*{\bibfont}{\footnotesize}
\printbibliography
\end{document}